\documentclass[british]{llncs}
\usepackage[latin9]{inputenc}
\usepackage{verbatim}
\usepackage{float}
\usepackage{amstext}
\usepackage{amssymb}
\usepackage{graphicx}
\usepackage{hyperref}
\hypersetup{pdftitle={Scalable Verification of MDPs}}

\makeatletter
 \usepackage[ruled,vlined]{algorithm2e}

\usepackage{tikz}

\makeatother

\usepackage{babel}
\begin{document}

\title{Scalable Verification of Markov Decision Processes}

\author{Axel Legay, Sean Sedwards and Louis-Marie Traonouez}

\institute{Inria Rennes -- Bretagne Atlantique}
\maketitle
\begin{abstract}
Markov decision processes (MDP) are useful to model concurrent process
optimisation problems, but verifying them with numerical methods is
often intractable. Existing approximative approaches do not scale
well and are limited to memoryless schedulers. Here we present the
basis of scalable verification for MDPSs, using an $\mathcal{O}(1)$
memory representation of history-dependent schedulers. We thus facilitate
scalable learning techniques and the use of massively parallel verification.
\end{abstract}

\section{Introduction}

Markov decision processes (MDP) describe systems that interleave nondeterministic
\emph{actions} and probabilistic transitions, possibly with\emph{
}rewards or costs assigned to the actions \cite{Bellman1957,Puterman1994}.
This model has proved useful in many real optimisation problems and
may also be used to represent concurrent probabilistic programs (see,
e.g., \cite{BiancoDeAlfaro1995,BaierKatoen2008}). Such models comprise
probabilistic subsystems whose transitions depend on the states of
the other subsystems, while the order in which concurrently enabled
transitions execute is nondeterministic. This order may radically
affect the expected reward or the probability that a system will satisfy
a given property. It is therefore useful to calculate the upper and
lower bounds of these quantities.

Fig. \ref{fig:MDP} shows a typical fragment of an MDP. Referring
in parentheses to the labels in the figure, the execution semantics
are as follows. In a given state ($s_{0}$), an action ($a_{1},a_{2},\dots$)
is chosen nondeterministically to select a distribution of probabilistic
transitions ($p_{1},p_{2},\dots$ or $p_{3},p_{4}$, etc.). A probabilistic
choice is then made to select the next state ($s_{1},s_{2},s_{3},s_{4},\dots$).
To each of the actions may be associated a reward ($r_{1},r_{2},\dots$),
allowing values to be assigned to sequences of actions.

\begin{figure}
\begin{minipage}[t]{0.4\columnwidth}%
\centering
\begin{tikzpicture}[every node/.style={draw,circle,inner sep=1pt}]
\draw(0,0)node(0){$s_0$}(-1,-1)node[fill=black](1){}
(0,-1)node[fill=black](2){}(-2,-2)node(3){$s_1$}(-1,-2)node(4){$s_2$}
(0,-2)node(5){$s_3$}(1,-2)node(6){$s_4$};
\draw[->,every node/.style={circle,inner sep=1pt}]
(0)edge[<-](0,0.75)
(0)edge node[left]{$r_1~$}node[above left]{$~a_1$}(1)
(0)edge node[left]{$a_2$}node[below left]{$r_2~$}(2)
(1)edge node[left]{$p_1$}(3)(1)edge node[below left]{$p_2$}(4)
(2)edgenode[below left]{$p_3$}(5)(2)edgenode[right]{$~p_4$}(6);
\draw[every edge/.style={dashed,draw}]
(3)edge(-2.25,-2.5)(3)edge(-1.75,-2.5)(4)edge(-1.25,-2.5)(4)edge(-0.75,-2.5)
(5)edge(-.25,-2.5)(5)edge(0.25,-2.5)(6)edge(0.75,-2.5)(6)edge(1.25,-2.5)(1)edge(-0.5,-1.5)(0)edge(1,-1);
\end{tikzpicture}\caption{Fragment of a typical Markov decision process.\label{fig:MDP}}
\end{minipage}\qquad{}%
\begin{minipage}[t]{0.53\columnwidth}%
\centering
\begin{tikzpicture}[every node/.style={circle,inner sep=1pt}]
\draw (0,0)node[draw](0){$s_0$}(0,-2)node[draw](3){$s_1$}(0)node[above right=0.5em]{$\models\neg\psi$}(3)node[rectangle,below=0.8em]{$\models\psi$};
\draw(0)edge[<-](0,0.75)(-1.3,-1)node[fill=black](2){}(1.3,-1)node[fill=black](4){}(0,-1)node(1)[fill=black]{}(3);
\draw[every edge/.style={->,bend right=39,draw}](0)edgenode[left]{$a_1$}(2)(2)edgenode[left]{$1-p_1$}(3)(4)edge[bend left=39]node[right]{$~1-p_2$}(3)(0)edge[bend left=39]node[right]{$a_2$}(4)(2)edge[bend right=20]node[above left]{$p_1\!\!$}(0)(4)edge[bend left=20]node[above right]{$p_2$}(0);
\draw[<-](1)edgenode[left]{$a_0$}(3)(0)edgenode[right]{1}(1);
\end{tikzpicture}\caption{MDP with different optima for general and memoryless schedulers when
$p_{1}\neq p_{2}$.\label{fig:historydependent}}
\end{minipage}
\end{figure}

To calculate the expected total reward or the expected probability
of a sequence of states, it is necessary to define how the nondeterminism
in the MDP will be resolved. In the literature this is often called
a\emph{ strategy}, a \emph{policy} or an \emph{adversary}. Here we
use the term \emph{scheduler} and focus on MDPs in the context of
\emph{model checking} concurrent probabilistic systems. Model checking
is an automatic technique to verify that a system satisfies a property
specified in temporal logic \cite{ClarkeEmersonAllenSifakis2009}.
\emph{Probabilistic} model checking quantifies the probability that
a probabilistic system will satisfy a property \cite{HanssonJonsson1994}.
Classic analysis of MDPs is concerned with finding schedulers that
maximise or minimise rewards \cite{Bellman1957,Puterman1994}. The
classic verification algorithms for MDPs are concerned with finding
schedulers that maximise or minimise the probability of a property,
or deciding the existence of a scheduler that ensures the probability
of a property is within some bound \cite{BiancoDeAlfaro1995}. Our
techniques can be easily extended to include rewards, but in this
work we focus on probabilities and leave rewards for future consideration.

\subsection{Schedulers and State Explosion\label{sec:schedulers}}

The classic algorithms to solve MDPs are \emph{policy iteration} and
\emph{value iteration} \cite{Puterman1994}. Model checking algorithms
for MDPs may use value iteration applied to probabilities \cite[Ch. 10]{BaierKatoen2008}
or solve the same problem using linear programming \cite{BiancoDeAlfaro1995}.
All consider \emph{history-dependent} schedulers. Given an MDP with
set of actions $A$, having a set of states $S$ that induces a set
of sequences of states $\Omega=S^{+}$, a history-dependent (general)
scheduler is a function $\mathfrak{S}:\Omega\rightarrow A$. A memoryless
scheduler is a function $\mathfrak{M}:S\rightarrow A$. Intuitively,
at each state in the course of an execution, a history-dependent scheduler
($\mathfrak{S}$) chooses an action based on the sequence of previous
states, while a memoryless scheduler ($\mathfrak{M}$) chooses an
action based only on the current state. History-dependent schedulers
therefore include memoryless schedulers.

Fig. \ref{fig:historydependent} illustrates a simple MDP for which
memoryless and history-dependent schedulers give different optima
for logical property $\mathbf{X}(\psi\wedge\mathbf{XG}^{t}\neg\psi)$
when $p_{1}\neq p_{2}$ and $t>0$. The property makes use of the
temporal operators \emph{next} ($\mathbf{X}$) and \emph{globally}
($\mathbf{G}$). Intuitively, the property states that on the next
step $\psi$ will be true and, on the step after that, $\neg\psi$
will be remain true for $t+1$ time steps. The property is satisfied
by the sequence of states $s_{0}s_{1}s_{0}s_{0}\cdots$. If $p_{1}>p_{2}$,
the maximum probability for $s_{0}s_{1}$ is achieved with action
$a_{2}$, while the maximum probability for $s_{0}s_{0}$ is achieved
with action $a_{1}$. Given that both transitions start in the same
state, a memoryless scheduler will not achieve the maximum probability
achievable with a history-dependent scheduler.

The principal challenge of finding optimal schedulers is what has
been described as the `curse of dimensionality' \cite{Bellman1957}
and the `state explosion problem' \cite{ClarkeEmersonAllenSifakis2009}:
the number of states of a system increases exponentially with respect
to the number of interacting components and state variables. This
phenomenon has led to the design of sampling algorithms that find
`near optimal' schedulers to maximise rewards in discounted MDPs.
Probably the best known is the Kearns algorithm \cite{KearnsMansourNg2002},
which we briefly review in Section \ref{sec:related}.

The state explosion problem of model checking applied to purely probabilistic
systems has been well addressed by \emph{statistical} model checking
(SMC) \cite{YounesSimmons2002}. SMC uses an executable model to approximate
the probability that a system satisfies a specified property by the
proportion of simulation traces that individually satisfy it. SMC
algorithms work by constructing an automaton to accept only traces
that satisfy the property. This automaton may then be used to estimate
the probability of the property or to decide an hypothesis about the
probability. Typically, the probability of property $\varphi$ is
estimated by $\frac{1}{N}\sum_{i=1}^{N}\mathbf{1}(\omega_{i}\models\varphi)$,
where $\omega_{1},\dots,\omega_{N}$ are $N$ independently generated
simulation traces and $\mathbf{1}(\cdot)$ is an indicator function
that corresponds to the output of the automaton: it returns $1$ if
the trace is accepted and $0$ if it is not. $N$ is chosen a priori
to give the required statistical confidence (e.g., using a Chernoff
bound \cite{Okamoto1958}, see Section \ref{sub:chernoff}). Sequential
hypothesis tests (e.g., Wald's sequential probability ratio test \cite{Wald1945},
see Section \ref{sub:wald}) do not define $N$ a priori, but generate
simulation traces until an hypothesis can be accepted or rejected
with specified confidence. The state space of the system is not constructed
explicitly--states are generated on the fly during simulation--hence
SMC is efficient for large, possibly infinite state, systems. Moreover,
since the simulations are required to be statistically independent,
SMC may be easily and efficiently divided on parallel computing architectures.

SMC cannot be applied to MDPs without first resolving the nondeterminism.
Since nondeterministic and probabilistic choices are interleaved in
an MDP, schedulers are typically of the same order of complexity as
the system as a whole and may be infinite. As a result, existing SMC
algorithms for MDPs consider only memoryless schedulers and have other
limitations (see Section \ref{sec:related}).

\subsection{Our Approach}

We have created memory-efficient techniques to facilitate Monte Carlo
verification of nondeterministic systems, \emph{without} storing schedulers
explicitly. In essence, the possibly infinite behaviour of schedulers
is fully specified \emph{implicitly} by the seed of a pseudo-random
number generator. Our techniques therefore require almost no additional
memory over standard SMC. In doing this, we are the first to provide
the basis for a complete lightweight statistical alternative to the
standard numerical verification algorithms for MDPs. A further contribution
is our derivation of the statistical confidence bounds necessary to
test multiple schedulers. These results suggest obvious solutions
to problems encountered with existing algorithms that rely on multiple
statistical tests (e.g., \cite{Henriques-et-al2012}).

In this work we demonstrate the core ideas of our approach with simple
SMC algorithms that repeatedly sample from scheduler space. Practical
implementations require more sophisticated algorithms that adopt ``smart
sampling'' (optimal use of simulation budget) and lightweight learning
techniques. Some of our results make use of these ideas, but a full
exposition is not possible here.

\section{Related Work\label{sec:related}}

\noindent The Kearns algorithm \cite{KearnsMansourNg2002} is the
classic `sparse sampling algorithm' for large, infinite horizon,
discounted MDPs. It constructs a `near optimal' scheduler piecewise,
by approximating the best action from a current state using a stochastic
depth-first search. Importantly, optimality is with respect to rewards,
not probability (as required by standard model checking tasks). The
algorithm can work with large, potentially infinite state MDPs because
it explores a probabilistically bounded search space. This, however,
is exponential in the discount. To find the action with the greatest
expected reward in the current state, the algorithm recursively estimates
the rewards of successive states, up to some maximum depth defined
by the discount and desired error. Actions are enumerated while probabilistic
choices are explored by sampling, with the number of samples set as
a parameter. The error is specified as a maximum difference between
consecutive estimates, allowing the discount to guarantee that the
algorithm will eventually terminate.

There have been several recent attempts to apply SMC to nondeterministc
models \cite{Bogdoll2011,LassaignePeyronnet2012,Henriques-et-al2012,HartmannsTimmer2013}.
In \cite{Bogdoll2011,HartmannsTimmer2013} the authors present on-the-fly
algorithms to remove `spurious' nondeterminism, so that standard
SMC may be used. This approach is limited to the class of models whose
nondeterminism does not affect the resulting probability of a property--scheduling
makes no difference. The algorithms therefore do not attempt to address
the standard MDP model checking problems related to finding optimal
schedulers.

In \cite{LassaignePeyronnet2012} the authors first find a memoryless
scheduler that is near optimal with respect to a reward scheme and
discount, using an adaptation of the Kearns algorithm. This induces
a Markov chain whose properties may be verified with standard SMC.
By storing and re-using information about visited states, the algorithm
improves on the performance of the Kearns algorithm, but is thus limited
to memoryless schedulers that fit into memory. The near optimality
of the induced Markov chain is with respect to rewards, not probability,
hence \cite{LassaignePeyronnet2012} does not address the standard
model checking problems of MDPs.

In \cite{Henriques-et-al2012} the authors present an SMC algorithm
to decide whether there exists a memoryless scheduler for a given
MDP, such that the probability of a property is above a given threshold.
The algorithm has an inner loop that generates candidate schedulers
by iteratively improving a probabilistic scheduler according to sample
traces that satisfy the property. The algorithm is limited to memoryless
schedulers because the improvement process counts state-action pairs.
The outer loop tests the candidate scheduler against the hypothesis
using SMC and is iterated until an example is found or sufficient
attempts have been made. The inner loop does not in general converge
to the true optimum, but the outer loop randomly explores local maxima.
This makes the number of samples used by the inner loop critical:
too many may significantly reduce the scope of the random exploration
and thus reduce the probability of finding the global optimum. A further
problem is that the repeated hypothesis tests of the outer loop will
eventually produce erroneous results. We address this phenomenon in
Section \ref{sec:confidence}.

\medskip{}

We conclude that (\emph{i}) no previous approach is able to provide
a complete set of SMC algorithms for MDPs, (\emph{ii}) no previous
SMC approach considers history-dependent schedulers and (\emph{iii})
no previous approach facilitates lightweight sampling from scheduler
space.

\section{Schedulers as Seeds of Random Number Generators\label{sec:seeds}}

Storing schedulers as explicit mappings does not scale, so we have
devised a way to represent schedulers using uniform pseudo-random
number generators (PRNG) that are initialised by a \emph{seed} and
iterated to generate the next pseudo-random value. In general, such
PRNGs aim to ensure that arbitrary subsets of sequences of iterates
are uniformly distributed and that consecutive iterates are statistically
independent. PRNGs are commonly used to implement the uniform probabilistic
scheduler, which chooses actions uniformly at random and thus explores
all possible combinations of nondeterministic choices. Executing such
an implementation twice with the same seed will produce identical
traces. Executing the implementation with a different seed will produce
an unrelated set of choices. Individual deterministic schedulers cannot
be identified, so it is not possible to estimate the probability of
a property under a specific scheduler.

An apparently plausible solution is to use independent PRNGs to resolve
nondeterministic and probabilistic choices. It is then possible to
generate multiple probabilistic simulation traces per scheduler by
keeping the seed of the PRNG for nondetermistic choices fixed while
choosing random seeds for a separate PRNG for probabilistic choices.
Unfortunately, the schedulers generated by this approach do not span
the full range of general or even memoryless schedulers. Since the
sequence of iterates from the PRNG used for nondeterministic choices
will be the same for all instantiations of the PRNG used for probabilistic
choices, the $i^{\mathrm{th}}$ iterate of the PRNG for nondeterministic
choices will always be the same, regardless of the state arrived at
by the previous probabilistic choices. The $i^{\mathrm{th}}$ chosen
action can be neither state nor trace dependent.

\subsection{General Schedulers Using Hash Functions\label{sec:hash}}

Our solution is to construct a per-step PRNG seed that is a \emph{hash}
of the an integer identifying a specific scheduler concatenated with
an integer representing the sequence of states up to the present.

We assume that a state of an MDP is an assignment of values to a vector
of system variables $v_{i},i\in\{1,\dots,n\}$. Each $v_{i}$ is represented
by a number of bits $b_{i}$, typically corresponding to a primitive
data type (\emph{int}, \emph{float}, \emph{double}, etc.). The state
can thus be represented by the concatenation of the bits of the system
variables, such that a sequence of states may be represented by the
concatenation of the bits of all the states. Without loss of generality,
we interpret such a sequence of states as an integer of $\sum_{i=1}^{n}b_{i}$
bits, denoted $\overline{s}$, and refer to this in general as the
\emph{trace vector}. A scheduler is denoted by an integer $\sigma$,
which is concatenated to $\overline{s}$ (denoted $\sigma:\overline{s}$)
to uniquely identify a trace and a scheduler. Our approach is to generate
a hash code $h=\mathcal{H}(\sigma:\overline{s})$ and to use $h$
as the seed of a PRNG that resolves the next nondeterministic choice.

The hash function $\mathcal{H}$ thus maps $\sigma:\overline{s}$
to a seed that is deterministically dependent on the trace and the
scheduler. The PRNG maps the seed to a value that is uniformly distributed
but nevertheless deterministically dependent on the trace and the
scheduler. In this way we approximate the scheduler functions $\mathfrak{S}$
and $\mathfrak{M}$ described in Section \ref{sec:schedulers}. Importantly,
our technique only relies on the standard properties of hash functions
and PRNGs. Algorithm \ref{alg:simulate} is the basic simulation function
of our algorithms.

\begin{algorithm}
\KwIn{\\\Indp

$\mathcal{M}$: an MDP with initial state $s_{0}$

$\varphi$: a property

$\sigma$: an integer identifying a scheduler

}\KwOut{\\\Indp

$\omega$: a simulation trace

}\BlankLine

Let $\mathcal{U}_{\mathrm{prob}},\mathcal{U}_{\mathrm{nondet}}$ be
uniform PRNGs with respective samples $r_{\mathrm{pr}},r_{\mathrm{\mathrm{nd}}}$

Let $\mathcal{H}$ be a hash function

Let $s$ denote a state, initialised $s\leftarrow s_{0}$

Let $\omega$ denote a trace, initialised $\omega\leftarrow s$

Let $\overline{s}$ be the trace vector, initially empty

Set seed of $\mathcal{U}_{\mathrm{prob}}$ randomly

\While{$\omega\models\varphi$ is not decided}{

$\overline{s}\leftarrow\overline{s}:s$ 

Set seed of $\mathcal{U}_{\mathrm{nondet}}$ to $\mathcal{H}(\sigma:\overline{s})$

Iterate $\mathcal{U}_{\mathrm{nondet}}$ to generate $r_{\mathrm{nd}}$
and use to resolve nondeterministic choice

Iterate $\mathcal{U}_{\mathrm{prob}}$ to generate $r_{\mathrm{pr}}$
and use to resolve probabilistic choice

Set $s$ to the next state

$\omega\leftarrow\omega:s$

}

\caption{Simulate\label{alg:simulate}}
\end{algorithm}

\vspace{-1.5em}

\subsection{An Efficient Iterative Hash Function}

To implement our approach, we have devised an efficient hash function
that constructs seeds incrementally. The function is based on modular
division \cite[Ch. 6]{Knuth1998}, such that $h=(\sigma:\overline{s})\bmod m$,
where $m$ is a suitably large prime.

Since $\overline{s}$ is a concatenation of states, it is usually
very much larger than the maximum size of integers supported as primitive
data types. Hence, to generate $h$ we use Horner's method \cite{Horner1819}\cite[Ch. 4]{Knuth1998}:
we set $h_{0}=\sigma$ and find $h\equiv h_{n}$ ($n$ as given in
Section \ref{sec:hash}) by iterating the recurrence relation 
\begin{equation}
h_{i}=(h_{i-1}2^{b_{i}}+v_{i})\bmod m.\label{eq:horner}
\end{equation}

The size of $m$ defines the maximum number of different hash codes.
The precise value of $m$ controls how the hash codes are distributed.
To avoid collisions, a simple heuristic is that $m$ should be a large
prime not close to a power of 2 \cite[Ch. 11]{CormenLeiersonRivestStein2009}.
Practically, it is an advantage to perform calculations using primitive
data types that are native to the computational platform, so the sum
in (\ref{eq:horner}) should be less than or equal to the maximum
permissible value. To achieve this, given $x,y,m\in\mathbb{N}$, we
note the following congruences:
\begin{eqnarray}
(x+y)\bmod m & \equiv & (x\bmod m+y\bmod m)\bmod m\label{eq:modadd}\\
(xy)\bmod m & \equiv & ((x\bmod m)(y\bmod m))\bmod m\label{eq:modmul}
\end{eqnarray}
The addition in (\ref{eq:horner}) can thus be re-written in the form
of (\ref{eq:modadd}), such that each term has a maximum value of
$m-1$:
\begin{equation}
h_{i}=((h_{i-1}2^{b_{i}})\bmod m+(v_{i})\bmod m)\bmod m\label{eq:horner2}
\end{equation}

To prevent overflow, $m$ must be no greater than half the maximum
possible integer. Re-writing the first term of (\ref{eq:horner2})
in the form of (\ref{eq:modmul}), we see that before taking the modulus
it will have a maximum value of $(m-1)^{2}$, which will exceed the
maximum possible integer. To avoid this, we take advantage of the
fact that $h_{i-1}$ is multiplied by a power of 2 and that $m$ has
been chosen to prevent overflow with addition. We thus apply the following
recurrence relation:

\begin{equation}
(h_{i-1}2^{j})\bmod m=(h_{i-1}2^{j-1})\bmod m+(h_{i-1}2^{j-1})\bmod m\label{eq:shiftmod}
\end{equation}

Equation (\ref{eq:shiftmod}) allows our hash function to be implemented
using efficient native arithmetic. Moreover, we infer from (\ref{eq:horner})
that to find the hash code corresponding to the current state in a
trace, we need only know the current state and the hash code from
the previous step. When considering memoryless schedulers we need
only know the current state.

\section{Confidence with Multiple Estimates\label{sec:confidence}}

The Chernoff bound \cite{Okamoto1958,Chernoff1952} and Wald sequential
probability ratio test \cite{Wald1945} are commonly used to bound
errors of SMC algorithms. Their guarantees are probabilistic, such
that with specified non-zero probability they produce an incorrect
result. If such bounds are used on $M$ schedulers, some of whose
true probabilities lie in the interval $(0,1)$, then as $M\rightarrow\infty$
the probability of encountering an error is a.s. $1$. In particular,
the maximum and minimum estimates will tend to 1 and 0, respectively,
regardless of the true values.

To overcome this phenomenon, in Sects. \ref{sub:wald} and \ref{sub:chernoff}
we derive new confidence bounds to allow SMC algorithms to test multiple
schedulers. We illustrate their use with simple algorithms that sample
$M$ schedulers at random, where $M$ is a parameter. These algorithms
are the basis of a technique we call ``smart sampling'', which can
exponentially improve convergence. The basic idea is to assign part
of the simulation budget to obtain a coarse estimate of the extremal
probabilities and to use this information to generate a set of schedulers
that contains a ``good'' scheduler with high probability. The remaining
budget is used to refine the set to find the best scheduler. Smart
sampling has provided improvements of several orders of magnitude
with the illustrated examples and is the subject of ongoing development.
Lack of space prevents further discussion.

\subsection{Sequential Probability Ratio Test for Multiple Schedulers\label{sub:wald}}

\noindent The sequential probability ratio test (SPRT) of Wald \cite{Wald1945}
evaluates hypotheses of the form $\mathrm{P}(\omega\models\varphi)\bowtie p$,
where $\bowtie\in\{\leq,\geq\}$. The SPRT distinguishes between two
hypotheses, $H_{0}:\mathrm{P}(\omega\models\varphi)\geq p^{0}$ and
$H_{1}:\mathrm{P}(\omega\models\varphi)\leq p^{1}$, where $p^{0}>p^{1}$.
Hence, to evaluate $\mathrm{P}(\omega\models\varphi)\bowtie p$, the
SPRT requires a region of indecision (an `indifference region' \cite{YounesSimmons2002})
which may be specified by parameter $\vartheta$, such that $p^{0}=p+\vartheta$
and $p^{1}=p-\vartheta$. The SPRT also requires parameters $\alpha$
and $\beta$, which specify the maximum acceptable probabilities of
errors of the first and second kind, respectively. An error of the
first kind is incorrectly rejecting a true $H_{0}$; an error of the
second kind is incorrectly accepting a false $H_{0}$. To choose between
$H_{0}$ and $H_{1}$, the SPRT defines the probability ratio 
\[
\mathit{ratio}=\prod_{i=1}^{n}\frac{(p^{1})^{\mathbf{1}(\omega_{i}\models\varphi)}(1-p^{1})^{\mathbf{1}(\omega_{i}\not\models\varphi)}}{(p^{0})^{\mathbf{1}(\omega_{i}\models\varphi)}(1-p^{0})^{\mathbf{1}(\omega_{i}\not\models\varphi)}},
\]
where $n$ is the number of simulation traces $\omega_{i}$, $i\in\{1,\dots$,
$n\}$, generated so far. The test proceeds by performing a simulation
and calculating $\mathit{ratio}$ until one of two conditions is satisfied:
$H_{1}$ is accepted if $\mathit{ratio}\geq(1-\beta)/\alpha$ and
$H_{0}$ is accepted if $\mathit{ratio}\leq\beta/(1-\alpha)$.

To decide whether there exists a scheduler such that $\mathrm{P}(\omega\models\varphi)\bowtie p$,
we would like to apply the SPRT to multiple (randomly chosen) schedulers.
The idea is to test different schedulers, up to some specified number
$M$, until an example is found. Since the probability of error with
the SPRT applied to multiple hypotheses is cumulative, we consider
the probability of no errors in any of $M$ tests. Hence, in order
to ensure overall error probabilities $\alpha$ and $\beta$, we adopt
$\alpha_{M}=1-\sqrt[M]{1-\alpha}$ and $\beta_{M}=1-\sqrt[M]{1-\beta}$
in our stopping conditions. $H_{1}$ is accepted if $\mathit{ratio}\geq(1-\beta_{M})/\alpha_{M}$
and $H_{0}$ is accepted if $\mathit{ratio}\leq\beta_{M}/(1-\alpha_{M})$.
Algorithm \ref{alg:SPRT} demonstrates the sequential hypothesis test
for multiple schedulers. If the algorithm finds an example, the hypothesis
is true with at least the specified confidence.%

\begin{algorithm}[h]
\KwIn{\\\Indp

$\mathcal{M},\varphi$: the MDP and property of interest

$H\in\{H_{0},H_{1}\}$: the hypothesis of interest with threshold
$p\pm\vartheta$

$\alpha,\beta$: the desired error probabilities of $H$

$M$: the maximum number of schedulers to test

}\KwOut{The result of the hypothesis test}\BlankLine

Let $p^{0}=p+\vartheta$ and $p^{1}=p-\vartheta$ be the bounds of
$H$

Let $\alpha_{M}=1-\sqrt[M]{1-\alpha}$ and $\beta_{M}=1-\sqrt[M]{1-\beta}$

Let $A=(1-\beta_{M})/\alpha_{M}$ and $B=\beta_{M}/(1-\alpha_{M})$

Let $\mathcal{U}_{\mathrm{seed}}$ be a uniform PRNG and $\sigma$
be its sample

\For{$i\in\{1,\dots,M\}$ while $H$ is not accepted}{

Iterate $\mathcal{U}_{\mathrm{seed}}$ to generate $\sigma_{i}$

Let $\mathit{ratio}=1$

\While{$\mathit{ratio}<A\wedge\mathit{ratio}>B$}{

$\omega\leftarrow\mathrm{Simulate}(\mathcal{M},\varphi,\sigma_{i})$

$\mathit{ratio}\leftarrow\frac{(p^{1})^{\mathbf{1}(\omega\models\varphi)}(1-p^{1})^{\mathbf{1}(\omega\not\models\varphi)}}{(p^{0})^{\mathbf{1}(\omega\models\varphi)}(1-p^{0})^{\mathbf{1}(\omega\not\models\varphi)}}\mathit{ratio}$

}\If{$\mathit{ratio}\geq A\wedge H=H_{0}\vee\mathit{ratio}\leq B\wedge H=H_{1}$}{accept
$H$

}}

\caption{Hypothesis testing with multiple schedulers\label{alg:SPRT}}
\end{algorithm}

\subsection{Chernoff Bound for Multiple Schedulers\label{sub:chernoff}}

Given that a system has true probability $p$ of satisfying a property,
the Chernoff bound ensures $\mathrm{P}(\mid\hat{p}-p\mid\geq\varepsilon)\leq\delta$,
i.e., that the estimate $\hat{p}$ will be outside the interval $[p-\varepsilon,p+\varepsilon]$
with probability less than or equal to $\delta$. Parameter $\delta$
is related to the number of simulations $N$ by $\delta=2e^{-2N\varepsilon^{2}}$
\cite{Okamoto1958}, giving
\begin{equation}
N=\left\lceil (\ln2-\ln\delta)/(2\varepsilon^{2})\right\rceil .\label{eq:NChernoff}
\end{equation}
The user specifies $\varepsilon$ and $\delta$ and the SMC algorithm
calculates $N$ to guarantee the estimate accordingly. Equation (\ref{eq:NChernoff})
is derived from equations
\begin{equation}
\mathrm{P}(\hat{p}-p\geq\varepsilon)\leq e^{-2N\varepsilon^{2}}\qquad\textnormal{and}\mathrm{\qquad P}(p-\hat{p}\geq\varepsilon)\leq e^{-2N\varepsilon^{2}},\label{eq:Chmax}
\end{equation}

\noindent giving $N=\left\lceil (\ln\delta)/(2\varepsilon^{2})\right\rceil $
to satisfy either inequality.

We consider the strategy of sampling $M$ schedulers to estimate the
optimum probability. We thus generate $M$ estimates $\{\hat{p}_{1},\dots,\hat{p}_{M}\}$
and take either the maximum ($\hat{p}_{\max}$) or minimum ($\hat{p}_{\min}$),
as required. To overcome the cumulative probability of error with
the standard Chernoff bound, we specify that \emph{all} estimates
$\hat{p}_{i}$ must be within $\varepsilon$ of their respective true
values $p_{i}$, ensuring that any $\hat{p}_{\min},\hat{p}_{\max}\in\{\hat{p}_{1},\dots,\hat{p}_{M}\}$
are within $\varepsilon$ of their true value. Given (\ref{eq:Chmax})
and the fact that all estimates $\hat{p}_{i}$ are statistically independent,
the probability that all estimates are less than their upper bound
is expressed by $\mathrm{P}(\bigwedge_{i=1}^{M}\hat{p}i-p_{i}\leq\varepsilon)\geq(1-e^{-2N\varepsilon^{2}})^{M}$.
Hence, $\mathrm{P}(\bigvee_{i=1}^{M}\hat{p}_{i}-p_{i}\geq\varepsilon)\leq1-(1-e^{-2N\varepsilon^{2}})^{M}$.
This leads to the following expression for $N$, given parameters
$M$, $\varepsilon$ and $\delta$:
\begin{equation}
N=\left\lceil -\ln\left(1-\sqrt[M]{1-\delta}\right)/2\varepsilon^{2}\right\rceil \label{eq:NMChernoff}
\end{equation}
Since the case for $p_{\min}$ is symmetrical, (\ref{eq:NMChernoff})
also ensures $\mathrm{P}(p_{\min}-\hat{p}_{\min}\geq\varepsilon)\leq\delta$.
Hence, to ensure the more usual conditions that $\mathrm{P}(\mid p_{\max}-\hat{p}_{\max}\mid\geq\varepsilon)\leq\delta$
and $\mathrm{P}(\mid p_{\min}-\hat{p}_{\min}\mid\geq\varepsilon)\leq\delta$,

\begin{equation}
N=\left\lceil \left(\ln2-\ln\left(1-\sqrt[M]{1-\delta}\right)\right)/(2\varepsilon^{2})\right\rceil .
\end{equation}
$N$ scales logarithmically with $M$ (e.g., for $\varepsilon=\delta=0.01$,
$N\approx\log_{1.0002}(M)+26472$), making it tractable to consider
many schedulers. Algorithm \ref{alg:palaiohora} is the resulting
extremal probability estimation algorithm for multiple schedulers.

\begin{algorithm}[H]
\KwIn{\\\Indp

$\mathcal{M},\varphi$: the MDP and property of interest

$\varepsilon,\delta$: the required confidence bound

$M$: the number of schedulers to test

}\KwOut{Extremal estimates $\hat{p}_{\min}$ and $\hat{p}_{\max}$}\BlankLine

Let $N=\left\lceil \ln(2/(1-\sqrt[M]{1-\delta}\,))/(2\varepsilon^{2})\right\rceil $
be the no. of simulations per scheduler

Let $\mathcal{U}_{\mathrm{seed}}$ be a uniform PRNG and $\sigma$
its sample

Initialise $\hat{p}_{\min}\leftarrow1$ and $\hat{p}_{\max}\leftarrow0$

Set seed of $\mathcal{U}_{\mathrm{seed}}$ randomly

\For{$i\in\{1,\dots,M\}$}{

Iterate $\mathcal{U}_{\mathrm{seed}}$ to generate $\sigma{}_{i}$

Let $\mathit{truecount}=0$ be the initial number of traces that satisfy
$\varphi$

\For{$j\in\{1,\dots,N\}$}{

$\omega_{j}\leftarrow\mathrm{Simulate}(\mathcal{M},\varphi,\sigma_{i})$

$\mathit{truecount}\leftarrow\mathit{truecount}+\mathbf{1}(\omega_{j}\models\varphi)$

}

Let $\hat{p}_{i}=\mathit{truecount}/N$

\If{$\hat{p}_{\max}<\hat{p}_{i}$}{$\hat{p}_{\max}=\hat{p}_{i}$

}\If{$\hat{p}_{i}>0\wedge\hat{p}_{\min}>\hat{p}_{i}$}{$\hat{p}_{\min}=\hat{p}_{i}$

}}

\If{$\hat{p}_{\max}=0$}{No schedulers were found to satisfy $\varphi$

}

\caption{Extremal probability estimation with multiple schedulers\label{alg:palaiohora}}
\end{algorithm}

\subsection{Experiments}

We implemented Algorithms \ref{alg:SPRT} and \ref{alg:palaiohora}
in our statistical model checking platform \textsc{Plasma} \cite{PLASMAproject}
and performed a number of experiments.

\begin{figure}
\begin{minipage}[t]{0.48\columnwidth}%
\begin{center}
\includegraphics[height=0.68\columnwidth]{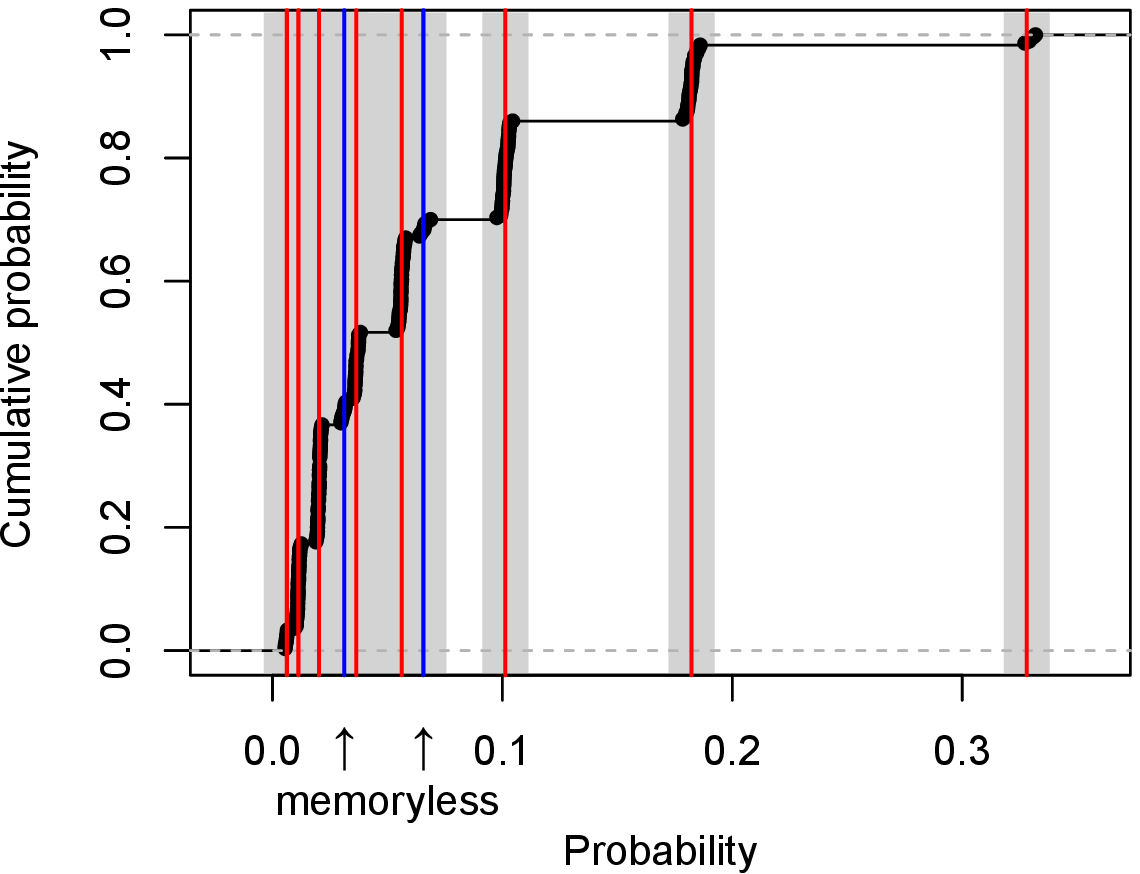}\caption{Empirical cumulative distribution of estimates from Algorithm \ref{alg:palaiohora}.\label{fig:schedulers}}

\par\end{center}%
\end{minipage}\hspace{1.5em}%
\begin{minipage}[t]{0.48\columnwidth}%
\begin{center}
\includegraphics[height=0.68\columnwidth]{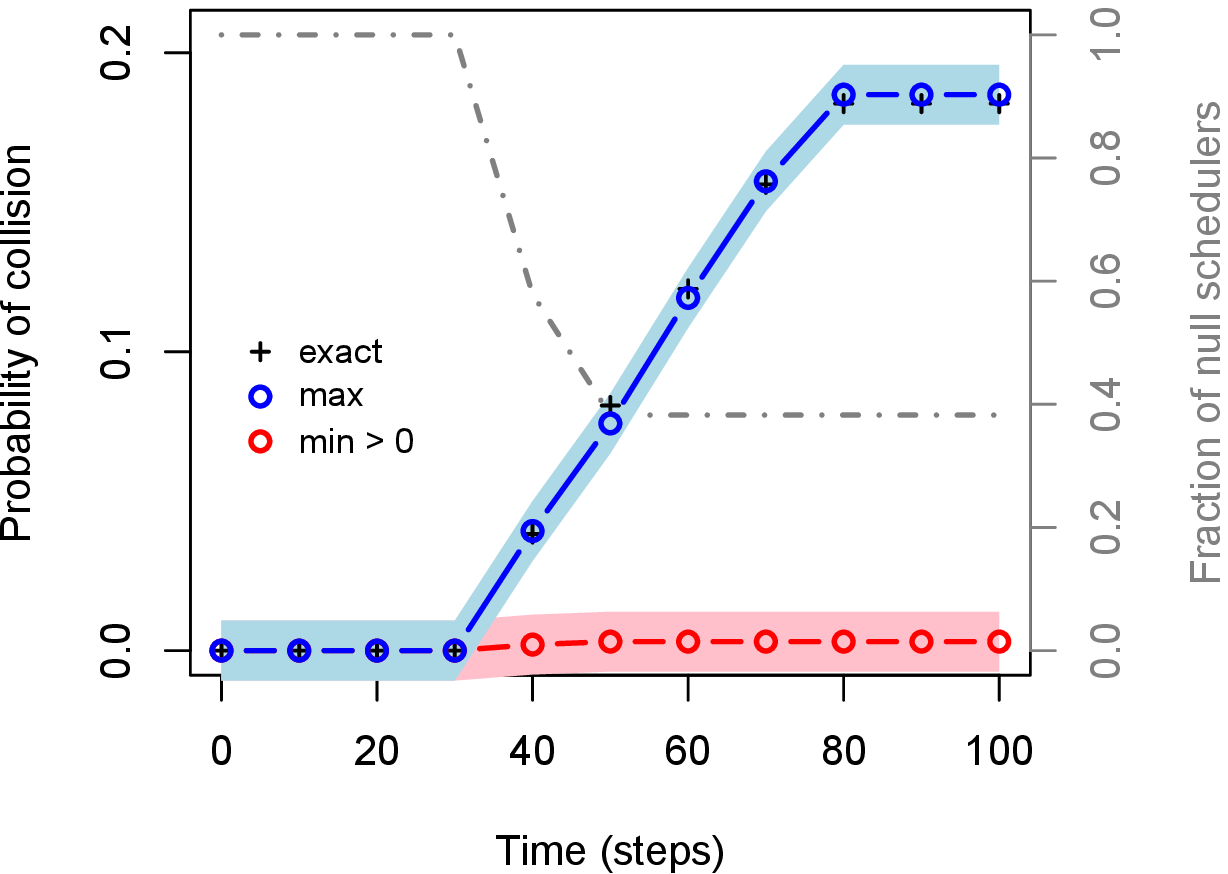}\caption{Max. and min. probabilities of second collision in WLAN protocol.\label{fig:WLAN}}

\par\end{center}%
\end{minipage}
\end{figure}

Figure \ref{fig:schedulers} shows the empirical cumulative distribution
of schedulers generated by Algorithm \ref{alg:palaiohora} applied
to the MDP of Fig. \ref{fig:historydependent}, using $p_{1}=0.9$,
$p_{2}=0.5$, $\varphi=\mathbf{X}(\psi\wedge\mathbf{XG}^{4}\neg\psi)$,
$\varepsilon=0.01$, $\delta=0.01$ and $M=300$. The vertical red
and blue lines mark the true probabilities of $\varphi$ under each
of the history-dependent and memoryless schedulers, respectively.
The grey rectangles show the $\pm\varepsilon$ error bounds, relative
to the true probabilities. There are multiple estimates per scheduler,
but all estimates are within their respective confidence bounds. Note
that the confidence is specified with respect to estimates, not with
respect to optimality. Defining confidence with respect to optimality
remains an open problem.

In Fig. \ref{fig:WLAN} we consider a reachability property of the
Wireless LAN (WLAN) protocol model of \cite{KwiatkowskaNormanSproston2002}.
The protocol aims to minimise ``collisions'' between devices sharing
a communication channel. We estimated the probability of the second
collision at time steps $\{0,10,\dots,100\}$, using Algorithm \ref{alg:palaiohora}
with $M=4000$ schedulers per point. Maximum and minimum estimated
probabilities are denoted by blue and red circles, respectively. Maximum
probabilities calculated by numerical model checking are denoted by
black crosses. The shaded areas indicate the $\pm\varepsilon$ error
of the estimates (Chernoff bound $\varepsilon=\delta=0.01$) and reveal
that our estimates are very close to the true values.

To demonstrate the scalability of our approach, we consider the choice
coordination model of \cite{NdukwuMcIver2010} and estimate the minimum
probability that a group of six tourists will meet within $T$ steps.
The model has a parameter ($\mathit{BOUND}$) that limits the state
space. We set $\mathit{BOUND}=100$, making the state space of $\approx5\times10^{16}$
intractable to numerical model checking. For $T=20$ and $T=25$ the
true minimum probabilities are respectively $0.5$ and $0.75$. Using
smart sampling and a Chernoff bound of $\varepsilon=\delta=0.01$,
we correctly estimate the probabilities to be $0.496$ and $0.745$
in a few tens of minutes on a standard laptop computer.

\section{Prospects and Challenges \label{sec:prospects}}

Our techniques are immediately extensible to continuous time MDPs
and other models that use nondeterminism. It is also seems simple
to consider MDPs with rewards. Although the presented algorithms are
not optimised with respect to simulation budget, in a forthcoming
work we introduce the notion of ``smart sampling'' to maximise the
chance of finding good schedulers with a finite budget.

A limitation of our approach is that the algorithms sample from only
a subset of possible schedulers. It is easy to construct examples
where good schedulers are vanishingly rare and will not be found.
Our ongoing focus is therefore to develop memory-efficient learning
techniques that construct schedulers piece-wise, to improve convergence
and consider a much larger set of schedulers.

\subsubsection*{Acknowledgement}

This work was partially supported by the European Union Seventh Framework
Programme under grant agreement no. 295261 (MEALS).

\bibliographystyle{abbrv}
\bibliography{FMDS2014}

\end{document}